\documentstyle[preprint,aps]{revtex}
\begin{document}
 \begin{title}
{\bf Weak-localization corrections to the conductivity \\of double quantum 
wells}
\end{title}
\author{ O. E. Raichev\cite{oleg} and P. Vasilopoulos\cite{takis} }
\address{\cite{takis}Concordia University, Department of Physics,\\ 1455 de 
Maisonneuve Blvd. Ouest,  Montr\'{e}al, Qu\'{e}bec, Canada, H3G 1M8  
\ \\
%\ \\
\cite{oleg}Institute of Semiconductor Physics, National Academy 
of Sciences of Ukraine,\\  Pr. Nauki 45, Kiev-28, 252650, Ukraine}
\date{\today}
\maketitle
\begin{abstract}

The weak-localization contribution $\delta\sigma(B)$ to the 
conductivity of a tunnel-coupled double-layer electron system is 
evaluated and its behavior in weak magnetic fields $B$ perpendicular 
or parallel to the layers is examined.  In a perpendicular field $B$, 
$\delta \sigma(B)$ increases and stays dependent on tunneling as 
long as the magnetic field is smaller than $\hbar/e D \tau_t$, where $D$ 
is the in-plane diffusion coefficient and $\tau_t$ the interlayer tunneling  
time. If $\tau_t$ is smaller than the inelastic scattering time, a parallel 
magnetic field also leads to a considerable increase of the
concuctivity starting with a $B^2$ law and saturating at fields higher 
than $\hbar/e Z \sqrt{D \tau_t}$, where $Z$ is the interlayer distance. 
In the limit of coherent tunneling, when $\tau_t$ is comparable to elastic 
scattering time, $\delta \sigma(B)$ differs from that of a single-layer 
system due to ensuing modifications of the diffusion coefficient. 
A possibility to probe the weak-localization effect in double-layer 
systems by the dependence of the conductivity on the gate-controlled 
level splitting is discussed.


\end{abstract}

\pagebreak

\section{Introduction}

The interference of electronic waves leads to negative corrections
to the conductivity of electron systems. This effect is known as weak
localization. It can be observed at very low temperatures, when the
inelastic scattering rate is so small that the phase coherence is kept
over many acts of elastic scattering. Although these quantum corrections
are usually small, they can be distinguished due to their specific dependence 
on  temperature, magnetic field,  and frequency of the applied electric field.

The fundamentals of the theory of weak localization have been developed
in Refs. 1-4 and a review is given in Ref. 5.  Recently considerable 
attention$^{6-16}$ has been drawn to  weak-localization phenomena 
in layered systems where characteristic features are caused 
by a dimensionality crossover. These studies have been applied mostly to 
multilayer systems (superlattices) although the cases of barrier-separated 
thin metallic films$^{9,10}$ and single-barrier structures$^{12}$ have also been 
 investigated. In this paper we calculate the weak-localization corrections 
to the in-plane conductivity of two tunnel-coupled two-dimensional (2D) layers, 
the system which is typically formed in double quantum well structures$^{17}$. 
Such double-layer electron systems represent an intermediate case between 
a single 2D layer and a three-dimensional superlattice and their various 
interesting properties are caused mainly by this fact. As concerns the weak 
localization problem, the tunnel coupling introduces an additional cutoff 
parameter in the diffusion pole, which implies that the weak-localization 
contribution to the conductivity depends on the strength of this coupling.  
This happens when the probability of  tunneling is not small in 
comparison to that of inelastic scattering. Another important 
consequence of the tunnel coupling is that a weak magnetic field $B$
applied {\it parallel } to the layers leads to a delocalization as 
does a field $B$ applied {\it perpendicular} to the layers$^4$. 
The physical origin of the effect of parallel field is explained 
in a similar way: this field introduces an additional phase for the 
electron which moves in one layer, then tunnels into another layer, 
moves there, and finally tunnels back and returns to the initial point.  
The described phenomena are the subject of  the present study.

The paper is organized as follows. In Sec. II we present the basic
formalism for the calculation of the weak-localization contribution to
the conductivity of a double-layer system. In Sec. III we calculate
the magnetoconductivity in a {\it perpendicular}  magnetic field $B$ and 
in Sec. IV we repeat the calculation for a {\it parallel } field $B$.  
In Sec. V we study the magnetoconductivity in conditions of coherent 
tunnel coupling, when the tunneling probability is comparable to or 
greater than the probability of elastic scattering. In Sec. VI we discuss 
the results and consider the possibility to probe the weak-localization
effects in double quantum wells by examining the dependence of the 
conductivity on the gate-controlled energy splitting of the lowest 
two levels of the double quantum well.

\section{Formalism}

The Hamiltonian of the double quantum well system is given, in the basis
of the left ($l$) and right ($r$) layer orbitals $\left| l \right>$ and $\left| 
r\right>$,  
by  

\begin{equation}
\hat{H}({\bf x})= \hat{P}_l \left[ E_l({\bf x}) + V_l({\bf x}) \right]
+ \hat{P}_r \left[ E_r({\bf x}) + V_r({\bf x}) \right] + \hat{h}
%1
\label{1}
\end{equation}
where ${\bf x}=(x,y)$ is the in-plane position vector; $E_j({\bf x}) = \left< j
\right| ( -i \hbar \partial/\partial {\bf x} + e{\bf A}({\bf x},z))^2
\left| j \right> /2m$ ($j=l,r$) is the matrix element of the kinetic 
energy operator, ${\bf A}({\bf x},z)$ the vector-potential,
$e$ the elementary charge, $m$ the effective mass, and $V_j({\bf x})= \left< 
j \right| V({\bf x},z) \left| j \right>$ the matrix element of the disorder 
potential.  Further, 

\begin{equation}
\hat{h}=\frac{\Delta}{2}\hat{\sigma}_z +T \hat{\sigma}_x ,
%2
\label{2}
\end{equation}
is the potential energy matrix of the double quantum well system
expressed through the energy level splitting $\Delta$ and tunneling
matrix element $T$.  Finally,  $\hat{\sigma}_i$ are the Pauli matrices and 
 $\hat{P}_l=(1+\hat{\sigma}_z)/2$ and $\hat{P}_r=(1-\hat{\sigma}_z)/2$  
the projection matrices. 

According to the Kubo formula, the zero-temperature dc conductivity is given by

\begin{equation}
\sigma_{\alpha \alpha'} = \frac{e^2 \hbar}{2 \pi m^2 L^2}
\sum_{b b'=R,A} (-1)^l  Tr \int \int d{\bf x} d{\bf x}'  %\nonumber \\
\left< \hat{\pi}^{\alpha}({\bf x})
\hat{G}_{\textstyle E_F}^{b}({\bf x},{\bf x}')
\hat{\pi}^{\alpha'}({\bf x}')
\hat{G}_{\textstyle E_F}^{b'}({\bf x}',{\bf x}) \right>
%3
\label{3}
\end{equation}
where $\hat{G}_{\textstyle E_F}^{R,A}({\bf x},{\bf x}')$ are
the Green's  functions of the electron system with  Fermi energy
$E_F$, $L^2$ is the normalization area, $\hat{\pi}^{\alpha}({\bf x})$ is 
the kinematic momentum operator, whose matrix elements are given as
$[\hat{ {\bf \pi}}({\bf x})]_{jj} = \left< j \right|-i \hbar
\partial/\partial {\bf x} + e{\bf A}({\bf x},z)  \left| j \right>$, 
and $\alpha$ and $\alpha'$ are the coordinate indices ($x$, $y$). The 
angular brackets $\left<...\right>$ denote statistical averaging, $Tr$ denotes
the  trace, and $l=1$ for $b=b'$ and $l=0$ for $b \neq b'$.

The weak-localization contribution $\delta \sigma_{\alpha \alpha'}$ to
the in-plane conductivity is given by the sum of an infinite set of diagrams 
with two or more maximally crossed impurity lines. Below we consider a system 
of randomly distributed elastic scatterers (impurities) described by the
correlation function $\left<V_j({\bf x}) V_{j'}({\bf x}') \right> = 
W_{jj'}({\bf x},{\bf x}')=w_{j} \delta_{jj'} \delta({\bf x} - {\bf x}')$. We 
obtain

\begin{eqnarray}
\nonumber \delta \sigma_{\alpha \alpha'}&=& \frac{e^2 \hbar}{2 \pi m^2 L^2}
\sum_{b b'=R,A} (-1)^l  \sum_{j j' j_1 j_2} \int \int d{\bf x} d{\bf x}'
\pi^{\alpha}_{jj}({\bf x})  G_{j j_1}^{b}({\bf x},{\bf x}_1)
G_{j_2 j'}^{b}({\bf x}_2,{\bf x}') \\*
\nonumber 
\ \\ 
&&\times \pi^{\alpha'}_{j'j'}({\bf x}')
G_{j' j_1}^{b'}({\bf x}',{\bf x}_1) G_{j_2 j}^{b'}({\bf x}_2,{\bf x})
\tilde{C}_{j_1 j_2}^{b b'}({\bf x}_1,{\bf x}_2),
%4
\label{4}
\end{eqnarray}
where $G_{j j'}^{b} ({\bf x},{\bf x}')=\left< [\hat{G}_{\textstyle
E_F}^{b}({\bf x},{\bf x}')]_{jj'} \right>$ are the averaged
one-particle Green's functions, and $\tilde{C}_{j_1 j_2}^{b b'}({\bf x}_1,
{\bf x}_2)$ is the Cooperon, the solution of the Bethe-Salpeter equation, given 
by

\begin{eqnarray}
\nonumber 
\tilde{C}_{j_1 j_2}^{bb'}({\bf x}_1,{\bf x}_2&)&= w_{j_1}
G_{j_1 j_2}^{b} ({\bf x}_1,{\bf x}_2) G_{j_1 j_2}^{b'}({\bf x}_1,
{\bf x}_2) w_{j_2} \\*
&&+ w_{j_1} \sum_{j} \int d {\bf x}  
G_{j_1 j}^{b} ({\bf x}_1,{\bf x}) G_{j_1 j}^{b'}({\bf x}_1,
{\bf x}) \tilde{C}_{j j_2}^{bb'} ({\bf x},{\bf x}_2).
%5
\label{5}
\end{eqnarray}
In the following we use Eqs. (\ref{4}) and (\ref{5}) to calculate $\delta
\sigma_{\alpha \alpha'}$ for both directions of the magnetic
field, perpendicular and parallel to the layers.

\section{Perpendicular magnetic field}

Consider the case when the  field $B$ is directed perpendicular to the
layers, ${\bf B}=(0,0,B)$. In the Landau gauge, ${\bf A}=(0,Bx,0)$, we can
write the Green's function as

\begin{eqnarray}
\nonumber 
G_{j_1 j_2}^{b} ({\bf x}_1,{\bf x}_2)&=& \frac{1}{2 \pi \ell^2}
e^{- i(x_1+x_2)(y_1-y_2)/2 \ell^2 } \ e^{ - ({\bf x}_1-{\bf x}_2)^2/4 \ell^2 }  
\\*
&&\times  
\sum_n L_n^0 \left(  ({\bf x}_1-
{\bf x}_2)^2/2 \ell^2 \right) G_{j_1 j_2}^{b}(n),
%6
\label{6}
\end{eqnarray}
where $G_{j_1 j_2}^{b}(n)$ is the averaged Green's function in the Landau-level
representation, $L_n^0$ are the Laguerre polynomials, and $\ell=
\sqrt{\hbar /eB}$  the magnetic length. From Eqs. (5) and (6) it
follows that the Cooperon may be written as

\begin{equation}
\tilde{C}_{j_1 j_2}^{bb'}({\bf x}_1,{\bf x}_2)= e^{ 
-i(x_1+x_2)(y_1-y_2)/\ell^2 } \  C_{j_1 j_2}^{bb'} (|{\bf x}_1-{\bf x}_2|),
%7
\label{7}
\end{equation}
where $C_{j_1 j_2}^{bb'}$ is its translationally 
invariant part, which is also called Cooperon in the following. 
Introducing $C_{j j'}^{bb'} (p)$ as the Fourier transform of 
$C_{j j'}^{bb'} (x)$, we obtain 

\begin{eqnarray}
\nonumber 
C_{j_1 j_2}^{bb'} (p)&=& w_{j_1} \Lambda_{j_1 j_2}^{bb'}(p) w_{j_2}
+ w_{j_1} \sum_{p', j} \int d {\bf x} e^{\frac{i}{\hbar} ({\bf p}'-
{\bf p}) \cdot {\bf x} } \\*
%\nonumber 
%\ \\
&&\times \Lambda_{j_1 j}^{bb'} \left( [(p'_x
- \hbar y/\ell^2)^2 + (p'_y + \hbar x/\ell^2)^2 ]^{1/2} \right) C_{j j_2}^{bb'} 
(p'),
%8
\label{8}
\end{eqnarray}
where $\Lambda_{j_1 j_2}^{bb'}(p)$ is the Fourier transform of the
translationally invariant part of the product $G_{j_1 j_2}^{b}
({\bf x}_1,{\bf x}_2) G_{j_1 j_2}^{b'} ({\bf x}_1, {\bf x}_2)$.
In the Landau-level representation it can be expressed as
  
\begin{equation}
\Lambda_{j j'}^{bb'}(p)= \frac{1}{2 \pi \ell^2}\sum_{nn'} (-1)^{n+n'}
\exp \left(-u \right)   
 L_n^{n-n'} \left( u \right)  L_{n'}^{n'-n}
\left( u \right) G_{j j'}^{b}(n) G_{j j'}^{b'}(n'),
%9
\label{9}
\end{equation}
where $u=p^2 \ell^2/2 \hbar^2$.  Below we consider very weak fields 
$B$, $\ell^2  \gg (v_{Fj}  \tau_j)^2$, where $v_{Fj}$ and $\tau_j$ are, 
respectively, 
the Fermi velocities and elastic scattering times in the layers. In this 
{\it diffusion limit} we can expand $\Lambda_{j_1 j}^{bb'} \left( 
\ldots\right)$ 
of Eq.  (\ref{8}) in series of $\hbar x/\ell^2$ and $\hbar y/\ell^2$.  
From Eq. (\ref{8}) we obtain a set of differential equations

\begin{equation}
C_{j_1 j_2}^{bb'} (p)= w_{j_1} \Lambda_{j_1 j_2}^{bb'}(p) w_{j_2}
+ w_{j_1}  %\left. 
\sum_{j} \left[ \Lambda_{j_1 j}^{bb'}(p)
%\nonumber \\
%\left. \times 
C_{j j_2}^{bb'} (p) -
\frac{\hbar^4}{2 \ell^4} \frac{d^2 \Lambda_{j_1 j}^{bb'}(p)}{ d p^2}
\frac{\partial^2}{\partial {\bf p}^2} C_{j j_2}^{bb'} (p) \right] .
%10
\label{10}
\end{equation}
In the diffusion limit we can also neglect the $B$
dependence of $\Lambda_{j_1 j_2}^{bb'}(p)$, using the expression

\begin{equation}
\Lambda_{j_1 j_2}^{bb'}(p)=\sum_{{\bf q}} G_{j_1 j_2}^{b}({\bf p}/2-{\bf q})
G_{j_1 j_2}^{b'}({\bf p}/2+{\bf q}),
%11
\label{11}
\end{equation}
where $G_{j_1 j_2}^{b}({\bf p})$ are the matrix elements of the
Green's function  

\begin{equation}
\hat{G}^{R,A}({\bf p}) = \left[E_F -\frac{p^2}{2m} - \hat{h}
\pm \hat{P}_l \frac{i\hbar}{2} \left( \frac{1}{\tau_l} + \frac{1}{\tau_{\varphi 
l}} \right)
\pm \hat{P}_r \frac{i\hbar}{2} \left( \frac{1}{\tau_r} + \frac{1}{\tau_{\varphi 
r}} \right) 
\right]^{-1}
%12
\label{12}
\end{equation}
describing the double-layer system in the absence of the magnetic field. 
The Fermi energy is measured from the center between the 
two levels of the double quantum well, the elastic scattering times 
are introduced as $\tau_j= \hbar^3/mw_j$, and $\tau_{\varphi j}$ are 
the phenomenologically introduced inelastic scattering times.

In the %is and next section we consider the 
weak coupling limit

\begin{equation}
\frac{1}{\tau_t}=\frac{2(T/\hbar)^2 \tau }{1+(\Delta \tau/ \hbar)^2 } \ll 
\frac{1}{\tau},
%13
\label{13}
\end{equation}
where $\tau= 2 \tau_l \tau_r/(\tau_l + \tau_r)$ is the average elastic 
scattering time and $\tau_t$ the tunneling time. A calculation 
of $\Lambda_{j_1 j_2}^{bb'}(p)$ in the limit (13) gives

\begin{equation}
\Lambda_{jj}^{RA}(p) \simeq
\frac{1}{w_j} \left( 1 -\frac{\tau_j}{\tau_{\varphi j}} - D_j \tau_j 
\left( {p\over\hbar} \right)^2 -
\frac{\tau_j}{\tau_t} \right) \\
\label{14}
\end{equation}
and

\begin{equation}
\Lambda_{lr}^{RA}(p)=\Lambda_{rl}^{RA}(p) \simeq \frac{\tau_l 
+\tau_r}{(w_l+w_r) 
\tau_t}\ .
%14,15
\label{15}
\end{equation}
In Eq. (14) we introduced the 2D diffusion coefficients in the layers 
$D_j=v_{Fj}^2 
\tau_j/2$. $\Lambda_{jj'}^{AR}(p)$ is given by Eqs.  (14) and (15) as well 
while $\Lambda_{jj'}^{RR}(p)$ and $\Lambda_{jj'}^{AA}(p)$ are small 
and can be neglected. A substitution of Eqs. (14) and (15) into Eq. (10) 
leads to the following equations

\begin{eqnarray}
- \frac{\hbar^2 D_l \tau_l}{\ell^4} \frac{\partial^2}{\partial {\bf p}^2}
C_{ll}^{RA}(p)  + \left(\frac{\tau_l}{\tau_{\varphi l}} + D_l \tau_l 
\left( {p\over\hbar} \right)^2
+ \frac{\tau_l}{\tau_t} \right) C_{ll}^{RA}(p)  - \frac{\tau_r}{\tau_t} 
C_{rl}^{RA}(p) = w_l ,\\
\nonumber
\ \\
- \frac{\hbar^2 D_r \tau_r}{\ell^4} \frac{\partial^2}{\partial {\bf p}^2}
C_{rl}^{RA}(p)  + \left(\frac{\tau_r}{\tau_{\varphi r}} + D_r \tau_r 
\left( {p\over\hbar} \right)^2
+ \frac{\tau_r}{\tau_t} \right) C_{rl}^{RA}(p)  - \frac{\tau_l}{\tau_t} 
C_{ll}^{RA}(p) = 0.
%16,17
\label{17}
\end{eqnarray}
$C_{rr}^{RA}(p)$ and $C_{lr}^{RA}(p)$  are given by Eqs. (16) and  (17)
with the indices $l$ and $r$  interchanged. 

Equations (16) and (17) can be diagonalized by the substitution 
$C_{ll}^{RA}(p)=C_{1}^l(p)+C_2^l(p)$ and $C_{rl}^{RA}(p)= 
\lambda_1 C_{1}^l(p)+ \lambda_2 C_2^l(p)$. For $C_k(p)$ ($k=$1,2) 
we obtain

\begin{eqnarray}
\left[- \frac{\hbar^2}{\ell^4} \frac{\partial^2}{\partial {\bf p}^2} + 
\left( {p\over\hbar} \right)^2 +q_k^2 \right]
C_k^l(p) = \frac{w_l}{\tau_l D_l} A_k, 
%18
\end{eqnarray}
where 
\begin{equation}
q_{1,2}^2=   \left(s_l + s_r \right)/2 \pm S/2, \ \ \ 
A_{1,2}=1/2  \pm (s_l -s_r)/2S.
%19 
\end{equation}
Here
$s_j=(D_j\tau_{\varphi j})^{-1} + (D_j\tau_t)^{-1}$, 
$s_t^2=(D_l D_r \tau_t^2)^{-1}$, and $S=[(s_l-s_r)^2+ 4 s_t^2]^{1/2}$.
Equation (18) is analogous to that for the Green's function of
a two-dimensional harmonic oscillator and its solution is obtained
in a straightforward way. The result is

\begin{equation}
C_{ll}^{RA}(p)= 2 \frac{w_l}{\tau_l D_l} \sum_n (-1)^n e^{ - u}
L_n^0 (2u)  %\nonumber \\
  \left[ \frac {A_1}{ (2/\ell^2) (2n+1) + q_1^2 } + 
\frac {A_2}{ (2/\ell^2) (2n+1) + q_2^2 } \right].
%20
\end{equation}
The expression for $C_{rr}^{RA}(p)$ is given by Eq.(20) with the layer 
indices interchanged.

Both terms in Eq. (20) contain diffusion poles modified due to tunneling. 
In the absence of tunneling (uncoupled layers) we obtain the well-known 
result$^4$ for each layer. Now we calculate the weak-localization contribution 
$\delta \sigma_{\alpha \alpha'}$. In the diffusion limit Eq. (4) gives

\begin{eqnarray}
\nonumber
\delta \sigma_{\alpha \alpha'} = \frac{e^2 \hbar}{2 \pi m^2 L^2}
\sum_{b b'=R,A} &(&-1)^l \sum_{j j' j_1 j_2} \sum_{p p'}
p_{\alpha} p'_{\alpha'} G_{j j_1}^{b}({\bf p}) G_{j_2 j'}^{b}({\bf 
p}')\\*
%\nonumber
%\ \\
&&\times
G_{j' j_1}^{b'}({\bf p}') G_{j_2 j}^{b'}({\bf p})
C_{j_1 j_2}^{b b'}(|{\bf p}+{\bf p}'|).
%21
\label{21}
\end{eqnarray}
In the limit (\ref{13}) we should retain  only the terms with $j=j'=j_1=j_2$.
The weak-localization contribution is expressed through $C_{ll}^{RA}(p)$ 
and $C_{rr}^{RA}(p)$. It gives a positive magnetoconductivity $\Delta 
\sigma = \delta \sigma(B)-\delta \sigma(0)$ as a sum of two parts$^9$

\begin{equation}
\frac{\Delta \sigma}{\delta \sigma_0} = f \left( 
\frac{4}{\ell^2 q_1^2}\right) + f \left( \frac{4}{\ell^2 q_2^2}\right). 
%22
\label{22}
\end{equation}
Here 
$\delta \sigma_0 =  e^2/2 \pi^2 \hbar$, and $f(x)=
\psi(1/2+1/x) +\ln x$, where $\psi(x)$ is the digamma function. 
In the absence of the magnetic field, the weak-localization 
contribution, within  logarithmic accuracy, is given as

\begin{equation}
\delta \sigma(0) = \delta \sigma_0
\ln \left( \frac{\tau_l \tau_r }{\tau_{\varphi l} \tau_{\varphi r}} + 
2 \ \frac{\tau_l \tau_r}{\tau_t \tau_{\varphi}} \right),
%23
\label{23}
\end{equation}
where we introduced an average inelastic scattering time $\tau_{\varphi}= 
2 \tau_{\varphi l} \tau_{\varphi r}/(\tau_{\varphi l} + 
\tau_{\varphi r} )$.  It is convenient to  analyze  
the results  
(22) and (23) in the symmetric case described by
$D_l=D_r=D$, $\tau_l=\tau_r$, and $\tau_{\varphi l}=\tau_{\varphi r}$. 
In this case Eq. (23) takes the form %becomes a sum of two contributions: 
$\delta \sigma(0)= \delta \sigma_0 [\ln(\tau/\tau_{\varphi}) + 
\ln(\tau/\tau_{\varphi} + 2\tau/\tau_t )]$. The first term is the 
same as in the absence of tunneling while the second 
one is tunneling-dependent. In the symmetric case 
Eq. (22)  gives

\begin{equation}
\frac{\Delta \sigma}{\delta \sigma_0}=
   f \left(\frac{B}{B_0} \right)
+ f \left(\frac{B}{B_0 (1+2 \tau_{\varphi}/\tau_t)} \right) ,
%24
\label{24}
\end{equation}
where $B_0= \hbar / 4e D \tau_{\varphi}$ is the characteristic 
field. In weak magnetic fields, the magnetoconductivity increases 
according to a $B^2$ law. In stronger fields, when the magnetic 
length $\ell$ becomes less than $\sqrt{D \tau_t}$, $\delta \sigma(B)$ 
loses its dependence on the tunneling and the system behaves 
like two decoupled layers. In the limit $B \gg B_0$, $B \gg B_0 
(\tau_{\varphi}/\tau_t)$ the magnetoconductivity is proportional 
to $2 \delta \sigma_0 \ln B$. For $\tau_t \ll \tau_{\varphi}$, an intermediate 
regime exists, $B_0  \ll B \ll  B_0 (\tau_{\varphi}/\tau_t)$, inwhich 
the magnetoconductivity is proportional to 
$\delta \sigma_0 \ln(B)$. Figure 1 shows the magnetic-field 
dependence of the magnetoconductivity described by Eq. (24) for 
several values of $\tau_{\varphi}/\tau_t$.

\section{Parallel magnetic field}

Due to the spatial separation of the layers, a   field $B$ applied
parallel to them can considerably influence the conductivity of the 
double-layer system, see, for example, Refs. 18-21. This effect occurs 
as a result of the modification of the electron energy spectrum 
of the coupled layers by the  field $B$. However, under the condition 
$\tau_t \gg \tau$ the coherent tunnel coupling is suppressed by 
the elastic scattering and the effect disappears. In contrast,  the 
influence of this in-plane field $B$ on the weak-localization part of 
the conductivity should exist as long as $\tau_t < \tau_{\varphi}$;
the corresponding characteristic field $B_1$ has to be much smaller. 
An assessment of  this influence is given below in the limit (13).

The parallel field renders the electron Green's function anisotropic. 
For ${\bf B}=(0,B,0)$ we choose ${\bf A}= (B(z-z_0),0,0)$ with
$z_0=(\left<l \right| z \left| l \right> + \left<r \right| z
\left| r \right>)/2$ and obtain$^{19,20}$

\begin{equation}
\hat{G}^{R,A}({\bf p}) = \left[E_F -\frac{p^2}{2m} -
\frac{\Delta_{p_x}}{2}\hat{\sigma}_z - T \hat{\sigma}_x
\pm \hat{P}_l \frac{i\hbar}{2} \left( \frac{1}{\tau_l} + \frac{1}{\tau_{\varphi 
l}} \right)
\pm \hat{P}_r \frac{i\hbar}{2} \left( \frac{1}{\tau_r} + \frac{1}{\tau_{\varphi 
r}} \right) 
\right]^{-1},
%25
\label{25}
\end{equation}
where $\Delta_{p_x}=\Delta - \omega_c Z p_x + \delta_B$, $\omega_c=eB/m$
is the cyclotron frequency, $Z=\left<r \right| z \left| r \right> -
\left<l \right| z \left| l \right>$ is the interlayer distance, and
$\delta_B \sim B^2$ is a small field-dependent correction to the level 
splitting which can be neglected for small $B$.

Using Eq. (\ref{25}) we can calculate $\Lambda_{jj'}^{bb'}({\bf p})$ from
Eq. (\ref{11}).  In contrast to the previous section, $\Lambda_{jj'}^{bb'}({\bf 
p})$ is now anisotropic. 
%due to the magnetic field-induced anisotropy of the Green function.
In the limit  $\ell^2 \gg Z v_{Fj} \tau_j$ instead of Eq. (14) we have

\begin{eqnarray}
\Lambda_{jj}^{RA}({\bf p}) \simeq
\frac{1}{w_j} \left( 1 -\frac{\tau_j}{\tau_{\varphi j}} - D_j \tau_j 
[(p_x \mp p_B)^2+p_y^2]/\hbar^2 - \frac{\tau_j}{\tau_t} \right), 
%26
\label{26}
\end{eqnarray}
where $p_B=m\omega_c Z= eBZ$, and the upper (lower) sign stands for $j=l$ 
($j=r$). 
Equation (\ref{15}) remains unchanged. The Bethe-Salpeter equation for the 
Cooperon now reduces to a set of linear algebraic equations

\begin{eqnarray}
C_{j_1 j_2}^{bb'} ({\bf p})= w_{j_1} \Lambda_{j_1 j_2}^{bb'}({\bf p})
w_{j_2} + w_{j_1} \sum_{j} \Lambda_{j_1 j}^{bb'}({\bf p})
C_{j j_2}^{bb'}({\bf p})
%29
\label{27}
\end{eqnarray}
whose solution is straightforward. A calculation of 
%the weak-localization  contribution 
$\delta \sigma_{\alpha \alpha'}$ can be done according to 
Eq. (\ref{21}), 
with $C_{j_1 j_2}^{b b'}(|{\bf p}+{\bf p}'|)$ replaced by $C_{j_1 j_2}^{bb'}
({\bf p}+{\bf p}')$ found from Eq. (27). Since $\tau_t \gg \tau$ holds, we 
again put $j=j'=j_1=j_2$ and finally obtain

\begin{eqnarray}
\Delta \sigma= \Delta \sigma_{sat} -  \delta \sigma_0  
\sum_{j=l,r} \int_0^{\infty} d x \frac{s_t^2}{(x+s_j)\sqrt{R_j(x)} } ~~, \\
\nonumber
\ \\
\Delta \sigma_{sat}=\delta \sigma_0   
\ln \left[ \left(1 +\frac{\tau_{\varphi l}}{\tau_t} \right) \left(1 +
\frac{\tau_{\varphi r}}{\tau_t} \right) \left/ \left(  1 + \frac{\tau_{\varphi 
l} + \tau_{\varphi r}}{\tau_t} \right) 
\right] \right.
%28,29
\end{eqnarray}
where $x=(p/\hbar)^2$ and $R_j(x)$ are the fourth-order
polynomials defined by

\begin{eqnarray}
\nonumber 
R_l(x)= x^4 + 2 (s_l+s_r - x_B ) x^3 + [(s_l+s_r)^2  %\nonumber \\
+ 2(s_l s_r - s_t^2)+ 2 x_B(s_r-2s_l) + x_B^2] x^2 \nonumber \\ 
+ 2[ (s_l+s_r)(s_l s_r - s_t^2) +x_B (2 s_l s_r -s_l^2 -s_t^2) %\nonumber \\ 
+  x_B^2 s_l] x+[s_l s_r -s_t^2 +x_B s_l]^2,
%30
\label{30}
\end{eqnarray}
with $x_B= (2 p_B/\hbar)^2$. The result for $R_r(x)$ is given by  Eq. (30)  
with the layer indices interchanged. Although the  field $B$ induces 
an anisotropy in the Green's functions, the localization correction $\delta 
\sigma(B)$ is isotropic because the main contribution in Eq. (21) comes 
from $|{\bf p}| \simeq |{\bf p}'| \simeq m v_{Fj}$. 

Equations (28)-(30) show that in the weak-field region $B \ll B_1$, where 
$B_1=\hbar/eZ \sqrt{D \tau_{\varphi}}$ and $D=2D_lD_r/(D_l+D_r)$ is 
the average diffusion coefficient, the magnetoconductivity follows a 
$B^2$ law:  

\begin{equation}
\frac{\Delta \sigma }{\delta \sigma_0} = \left( \frac{2 s_t e Z B}{\hbar S} 
\right)^2  \left[ \frac{s_l+s_r}{s_l s_r-s_t^2} -\frac{2}{S} \ln \left( 
\frac{s_l+s_r+
S}{s_l+s_r -S} \right) \right].
%31
\end{equation}
At $\tau_t \ll \tau_{\varphi j}$ we obtain $\Delta \sigma /\delta \sigma_0 = 
(B/B_1 )^2$. In the opposite limit, a simple expression for the 
magnetoconductivity can be obtained in the symmetric case,
$\Delta \sigma / \delta \sigma_0 = \frac{4}{3}  (\tau_{\varphi}/\tau_t)^2 
(B/B_1)^2$; it shows a substantial suppression of the magnetoconductivity  
as a result of decreasing tunneling probability. 

With the increase of $B$, the $B$-dependence becomes weaker.  
For $B \gg B_1$ and $B^2 \gg B_1^2 \tau_{\varphi} / \tau_t$, the 
integral in Eq. (28) can be neglected
%goes to zero, vanishes 
and the magnetoconductivity is saturated and equal to 
$\Delta \sigma_{sat}$ defined by Eq. (29). For $\tau_{\varphi}/\tau_t 
\ll 1$ the saturated value goes to zero. Recently, a saturation 
behavior of the weak-localization contribution in parallel magnetic 
fields has been theoretically found for superlattices$^{15}$. 

When the condition $\tau_t \ll \tau_{\varphi}$ holds, one can analytically
evaluate the magnetoconductivity for $B^2 \ll B_1^2 \tau_{\varphi}/ \tau_t$; 
the result is
\begin{equation}
\Delta \sigma = \delta \sigma_0 \ln [1+(B/B_1)^2].
%32
\label{32}
\end{equation}
It shows that in the intermediate region $B_1^2 \ll B^2 \ll B_1^2 
\tau_{\varphi}/\tau_t$, the magnetoconductivity follows a logarithmic law. 
In Fig. 2 we demonstrate a transition from a $B^2$-behavior at weak 
fields to the saturation at stronger fields for several values of the 
ratio $\tau_t/\tau_{\varphi}$. 

The tunneling rate $1/\tau_t$ can be varied by applying a transverse 
voltage which changes the level splitting $\Delta^{17}$. 
%which is easily controllable in experiments on double quantum wells$^{17}$ 
%by applying a transverse voltage. 
In the case under consideration 
$E_F \gg \hbar/\tau$, a relatively small variation of $\Delta$ has no 
considerable influence on the electron densities in the wells but can 
dramatically modify the tunneling rate. Fig. 3 shows the $\Delta$-
dependence of $\delta \sigma(B)- \delta \sigma(0)|_{\Delta=0}$ for 
several values of the magnetic field, for both {\it parallel }(solid) and 
{\it perpendicular} (dashed lines) fields; the thick solid line is the 
result for $B=0$. The curves have a typical resonanse-like behavior 
that is affected differently by the two field orientations. An increase 
of $\Delta$ always leads to a decrease of the conductivity, because
in this way  the tunneling is suppressed and, therefore,  the 
interference is increased. The perpendicular magnetic field tends to smoothen 
this effect because it suppresses the dependence of the 
weak-localization part of the conductivity on tunneling, see Sec. III. On 
the other hand, the parallel  field tends to strengthen 
the tunneling dependence of the weak-localization part: in the saturation 
regime, cf. Fig. 2, this dependence is the strongest. 

Above we calculated the magnetoconductivity in the weak-tunneling 
regime. Below we evaluate it again in the regime of coherent tunnel coupling. 

\section{Coherent tunneling regime}

The aim of this section is to study the weak localization effects at 
sufficiently strong, coherent tunnel coupling, described by
$\tau_t  \sim  \tau$ 
or even $\tau_t  \ll \tau$. In these conditions one always has $\tau_t 
\ll \tau_{\varphi}$, which means that there is no competition between 
the tunneling and inelastic scattering processes. As a result, 
instead of two diffusion poles, see Eq. (20), the expression for the Cooperon 
contains just one pole as for a single-layer system. Hovewer, 
the properties of this pole are somewhat different than those in 
case of a single 2D layer and require a separate consideration. 

Below we assume that $E_F$ is large in comparison with 
$T$ and $\Delta$. Introducing a single Fermi velocity $v_F$ for 
both layers, we write average diffusion coefficient as 
$D=v_F^2 \tau/2$. The calculation of the Cooperon according to 
Eqs. (11), (25), and (27) gives
\begin{equation}
C_{jj'}({\bf p}) =\frac{w_j w_{j'}}{w_l+w_r} \left[ \frac{\tau}{
\tau_{\varphi}} + \frac{D(\Delta) \tau}{\hbar^2} \left( p^2 + r(\Delta) 
p_B^2+ 2 \mu r(\Delta) p_B p_x \right) \right]^{-1}. 
\label{33}
\end{equation}
The inelastic scattering time enters the diffusion pole only as an 
average. On the other hand, the Cooperon depends on the asymmetry 
of the elastic scattering  described by the
parameter $\mu=(\tau_r-\tau_l)/(\tau_r+\tau_l)$. In Eq. (33) 
we have also introduced a $\Delta$-dependent diffusion coefficient 
according to
\begin{eqnarray}
D(\Delta)= \frac{D}{1-\mu^2 r(\Delta)},~~~ 
r(\Delta)=\frac{\Delta^2+(\hbar/\tau)^2 }{\Delta^2+ 4T^2 + (\hbar/\tau)^2 }.
%34
\end{eqnarray}
Note that $D(\Delta)$ differs from $D$ only if the elastic scattering 
is asymmetric.

If there is no parallel magnetic field ($p_B=0$), Eq. (33) has the same form as 
the Cooperon of a single-layer system. Therefore, in the diffusion regime 
Eq. (33) with a substitution $p^2 \rightarrow  4 \hbar e B (n+1/2)$ can be 
directly used to calculate the magnetconductivity in a perpendicular 
magnetic field $B$. Using Eq. (21), we obtain, in analogy with the 
single-layer case,
\begin{equation}
\Delta \sigma= \delta \sigma_0 
f \left( \frac{4 e D (\Delta) B \tau_{\varphi}}{\hbar} \right),
%35
\end{equation}
and, within the logarithmic accuracy,  $\delta \sigma(0)= \delta \sigma_0 
\ln( \tau/\tau_{\varphi})$. The 
only difference between Eq. (35) and the well-known expression for a 
single-layer 2D system is the replacement of $D$ in the latter by 
$D(\Delta)$. The physical meaning of this change is clear and
easily understood in terms of the resistance resonance phenomenon
$^{22,23}$, see also Ref. 17. In the corresponding theory$^{22,23}$ 
the factor $1/[1-\mu^2 r(\Delta)]$ describes the $\Delta$-dependence of 
the conductivity and of the diffusion coefficient $D$ of the double quantum 
wells with asymmetric scattering. With the increase of $|\Delta|$, 
$D$ increases and leads to an increase of the 
magnetoconductivity $\Delta \sigma$. 

Now we consider the case of a parallel magnetic field. Using Eqs. (33) 
and (21), we obtain   
\begin{equation}
\Delta \sigma = \delta \sigma_0 \ln \left[ 1 + r(\Delta) (B/B_1)^2 \right].
%36
\end{equation}
The magnetoconductivity described by Eq. (36) does not depend on 
the scattering asymmetry. On the other hand, its dependence on 
$\Delta$ has the same sign as in case of perpendicular field: when 
$\Delta$ increases, $r(\Delta)$ goes to unity and $\Delta \sigma$ 
increases. The behavior of the magnetoconductivity described by Eqs. (35) 
and (36) is illustrated in Fig. 4.

The results for magnetoconductivity described in this section are 
valid for $\tau_t \ll \tau_{\varphi}$ under conditions $D \tau_t e B/ \hbar \ll 
1$ 
for perpendicular field $B$ and $D \tau_t (e Z B/ \hbar )^2 \ll 1$ for parallel 
field $B$. In the limit (13), when $r(\Delta)=1$, Eq. (35) is equivalent to 
Eq. (22) written at $\tau_t \ll \tau_{\varphi}$ and $D \tau_t e B/ \hbar \ll 1$,
while Eq. (36) is equivalent to Eq. (32). 

\section{Discussion}

We have investigated the influence of the tunnel coupling betwen two 
2D layers on the weak-localization-induced magnetoconductivity. 
This couplng intruduces an extra degree of freedom for an electron 
giving it the possibility to tunnel between the layers and, therefore, it 
reduces the interference effects.  As a result, the weak-localization 
contribution is reduced as the tunneling rate $1/\tau_t$ overcomes 
the inelastic scattering rate. In particular, it leads to the weakening 
of the magnetoresistance in the perpendicular magnetic field, see 
Sec. III.

The results described in Sec. III are similar to those obtained 
in  Ref. $9$ for two thin metallic films divided by a barrier if 
one equals $\tau_t$ to the tunneling times $\tau_{12}$ and 
$\tau_{21}$ of Ref. 9. This is not surprising. When the inelastic 
scattering lengths are small in comparison to the film widths, the 
diffusion of the electrons in such a system proceeds in the same 
way as in coupled 2D layers. A characteristic feature of the 
double quantum well case, as opposed to the case of thin films, 
is the strong resonance dependence 
of the tunneling time $\tau_t$, cf.  Eq. (\ref{13}), on the energy-level 
splitting $\Delta$. The latter can be easily controlled by external 
gates enclosing the double quantum wells. This opens a new way 
to examine the weak-localization phenomena by  measuring the 
$\Delta$-dependence of the conductivity, see, for example, Fig. 3, 
in addition to  the $B$ dependence. The conductivity, due to its 
weak-localization part, is maximal when the tunneling 
resonance condition $\Delta=$0 holds (this can be called  
"delocalisation resonance"), while an increase of $\Delta$ 
increases the interference and leads to a decrease of the 
conductivity down to its value for uncoupled layers. On the 
other hand, if the scattering times in the layers are different, the 
main part of the conductivity will show the opposite effect (the 
resistance resonance$^{17}$). Although the relative size of the 
resistance resonance would be small due to the parameter 
$\tau/\tau_t$, it can obscure the delocalization resonance. Therefore,
the samples with symmetric scattering are preferrable for probing the 
interference corrections.   

In Sec. IV we studied the magnetoconcuctivity in a parallel magnetic field.
Since this magnetoconcuctivity can exist only in the presence of 
the tunnel coupling, it becomes more pronounced as the tunneling rate 
overcomes the inelastic scattering rate. Although at very small $B$ 
both parallel and perpendicular fields give $B^2$-corrections to the
magnetoconductivity, at higher fields the behavior becomes completely 
different. When the {\it perpendicular} field is larger than both $B_0$ and 
$B_0 \tau_{\varphi}/ \tau_t$, the weak-localization contribution 
logarithmically increases with $B$ and does not depend on tunnneling.  
When the {\it parallel } field is larger than both $B_1$ and $B_1 
\sqrt{\tau_{\varphi}/ \tau_t}$, $\delta \sigma(B)$ becomes saturated, 
i.e., it loses its dependence on $B$ but not that on tunneling. 
Since $(v_F/ Z) \sqrt{\tau_{\varphi} \tau} \gg 1$ holds, the characteristic 
parallel field $B_1$ is much larger than the characteristic perpendicular 
field $B_0$. 

With an increase of $\Delta$ both $\delta \sigma(B)$ and $\Delta \sigma$ 
in a parallel magnetic field decrease. However, in the case of coherent 
tunnel coupling (Sec. V) the behavior of the magnetoconductivity $\Delta \sigma$ 
versus $\Delta$ is opposite: an increase of $\Delta$ increases $\Delta \sigma$.
A transition between these two kinds of behavior can be explained 
as follows. When the tunnel coupling becomes so strong that 
$D \tau_t (e B Z/ \hbar)^2 \ll 1$, a change of $\tau_t$ with $\Delta$ 
no longer modifies the magnetic-field suppression of the interference, 
see Sec. IV. With a further increase of the tunneling rate, when 
the latter overcomes the elastic scattering rate, coherent 
coupling effects become important. Instead of the states localized in 
the wells, one has a pair of hybridized states, which are well-defined 
in the limit of $2T \gg \hbar/\tau$. At $\Delta=0$ the electron density 
in both these states is distributed equally among the wells, and the 
parallel magnetic field cannot reduce the interference in the way 
described in the Introduction. The magnetoconductivity is zero, which 
is given also by Eq. (36) at $2T \gg \hbar/\tau$ and $\Delta=0$. 
At $\Delta \neq 0$ the symmetry is broken. The two states are mostly
localized in different wells and the parallel magnetic 
field can reduce the interference because the transitions of electrons 
between these states are possible. In conclusion, the magnetoconductivity
increases with an increase of $\Delta$. 
 
Let us briefly discuss the main approximations used in this paper. Since we 
neglected the electron-electron interaction, which also leads to quantum 
corrections to the conductivity, the theory is applicable at  
temperatures considerably lower than $\hbar/\tau_{\varphi}$$^{24}$. 
We also neglected the possibility of spin-flip scattering, assuming that the 
relevant scattering time is larger than $\tau_{\varphi}$. The assumption of 
the $\delta$-correllated scattering does not lead to principal changes. 
For example, if one takes into account the interlayer correlation of the 
scattering amplitudes, one has  to replace only $1/\tau$ in Eq. (13) by $1/\tau 
- 1/\tau_{lr}$, where $ \tau_{lr}=\hbar^3/mw_{lr}$.

Throughout the paper we calculated the total, i.e. summed over both layers, 
in-plane conductivity of the double-layer system. Experimentally it corresponds 
to the double quantum wells with common contacts to the layers. The case 
of independently contacted layers does not require a separate consideration 
for the following reason. Since the system is assumed macroscopic, its in-plane 
size should substantially exceed the inelastic scattering length $v_F 
\tau_{\varphi}$. 
On the other hand, the tunnel coupling manifests itself in  
weak-localization 
phenomena under the condition $\tau_{\varphi} > \tau_t$. In such a case 
a characteristic current redistribution length$^{25}$ $l_s \sim \sqrt{v_F^2 
\tau \tau_t}$ is small in comparison to the size of the system, which 
means that for any case of independent contacting the double-layer 
system would behave as a system with common contacts.

\acknowledgements
This work was supported by the Canadian NSERC Grant No. OGP0121756. O. E. R. 
also acknowledges support from le Minist\'{e}re de l' Education du Qu\'{e}bec.

\clearpage 
%{\center FIGURE CAPTIONS\\}

\begin{figure}
\caption{ Magnetoconductivity of the double-layer system in a {\it 
perpendicular }
magnetic field $B$ for symmetric conditions. The curves are marked with the 
values of  the ratio $\tau_t/\tau_{\varphi}$  and $B_0=\hbar / 4 e D 
\tau_{\varphi}$}
\label{fig.1}
\ \\
\caption{Magnetoconductivity of the double-layer system in a {\it 
parallel } magnetic  field $B$ for symmetric conditions. The curves are marked 
as in Fig. 1  and $B_1=\hbar /e Z \sqrt{ D \tau_{\varphi}}$ }
\label{fig.2}
\ \\

\caption{Relative weak-localization correction to the conductivity of the 
double-layer system as a function of the level splitting  $\Delta$ for several 
values of the {\it parallel } (solid) and {\it perpendicular }(dashed) magnetic 
field. It is assumed that $\tau_t/\tau_{\varphi}=0.1$ at $\Delta=0$. The numbers 
next to the curves show the values of $B/B_0$ (perpendicular field) and 
$(B/B_1)^2$ (parallel field). }
\label{fig.3}

\ \\
\caption{Magnetoconductivity of the double-layer system in a coherent 
tunneling regime for $2T \tau/\hbar =4$ and $\mu=0.7$. The solid and dashed 
lines correspond to parallel and perpendicular field, respectively. The 
curves are marked with the values of  the ratio $\Delta/2T$.}
\label{fig.4}

\end{figure}

\clearpage 
\begin{references}
\bibitem[\star]{oleg}  E-mail: raichev@boltzmann.concordia.ca 
\bibitem[\diamond]{takis}  E-mail: takis@boltzmann.concordia.ca 

\bibitem{1} E. Abrahams, P. W. Anderson, D. C. Licciardello, and T. V. 
Ramakrishnan,
Phys. Rev. Lett. {\bf 42}, 637 (1979).

\bibitem{2} L. P. Gor'kov, D. Chmelnitski, and A. I. Larkin, Pisma ZhETP {\bf 
30}, 248
(1979).

\bibitem{3} S. Hikami, A. I. Larkin, and Y. Nagaoka, Progr. Theor. Phys. {\bf 
63}, 707
(1980).

\bibitem{4} B. L. Altshuler, D. Chmel'nitskii, A. I. Larkin, and P. A. Lee. 
Phys. Rev. B {\bf 22}, 5142 (1980).

\bibitem{5} P. A. Lee and T. V. Ramakrishnan, Rev. Mod. Phys. {\bf 57}, 287 
(1985); B. L. Altshuler and A. G. Aronov, in {\em Electron-electron 
interactions 
in disordered systems}, Elsevier Science Publishing, 1985, p.1; H. Fukuyama,
ibid., p.155.

\bibitem{6} E. P. Nakamedov, V. N. Prigodin, and Yu. A. Firsov, JETP Lett. {\bf 
43}, 743 
(1986).

\bibitem{7} W. Szott, C. Jedrzejek, and W. P. Kirk, Phys. Rev. B {\bf 40}, 1790 
(1989).

\bibitem{8} W. Szott, C. Jedrzejek, and W. P. Kirk, Superlattices and 
Microstructures
{\bf 11}, 199 (1992).

\bibitem{9} G. Bergmann, Phys. Rev. B {\bf 39}, 11280 (1989).

\bibitem{10} W. Wei and G. Bergmann, Phys. Rev. B {\bf 40}, 3364 (1989).

\bibitem{11} X. J. Lu and N. J. M.Horing, Phys. Rev. B {\bf 44}, 5651 (1991).

\bibitem{12} I. V. Lerner and M. E. Raikh, Phys. Rev. B {\bf 45}, 14036 (1991).

\bibitem{13} N. Dupuis and G. Montambaux, Phys. Rev. B {\bf 46}, 9603 (1992).

\bibitem{14} V. V. Dorin, Phys. Lett. A {\bf 183}, 233 (1993).

\bibitem{15} C. Mauz, A. Rosch, and P. Wolfe, Phys. Rev. B {\bf 56}, 10953 
(1997).

\bibitem{16} J. J. Mares, J. Kristofik, P. Bubik, E. Bulicius, K. Melichar, J. 
Pangrac, J. Novak, and S. Basenohrl, Phys. Rev. Lett {\bf 80}, 4020 (1998).

\bibitem{17} See, for example, N. K. Patel, A.  Kurobe, I. M. Castleton, E. 
B. Linfield, K. M. Brown, M. P. Grimshaw, D. A.  Ritchie, G. A. C. Jones, and 
M. Pepper, Semicond. Sci. Technol. {\bf 11}, 703 (1996).

\bibitem{18} J. A. Simmons, S. K. Lyo, N. E. Barff, and J. F. Klem, Phys. Rev. 
Lett. {\bf 73}, 2256 (1994).

\bibitem{19} Y. Berk, A. Kamenev, A. Palevski, L. N. Pfeiffer, and K. W. West,
Phys. Rev. B {\bf 51}, 2604 (1995).

\bibitem{20} F. T. Vasko and O. E. Raichev, Phys. Rev. B {\bf 52}, 16349 
(1995);O. E. Raichev and F. T. Vasko. Phys. Rev. B {\bf 53}, 1522 (1996).

\bibitem{21} T. Jungwirth, T. S. Lay, M. Smrcka, and M. Shayegan, Phys. Rev. B
{\bf 56}, 1029 (1997).

\bibitem{22} A. Palevski, F. Beltram, F. Capasso, L. N. Pfeiffer, and K. W. 
West, Phys. Rev. Lett. {\bf 65}, 1929 (1990).

\bibitem{23} F. T. Vasko, Phys. Rev. B. {\bf 47}, 2410 (1993).

\bibitem{24} B. L. Altshuler, A. G. Aronov, A. I. Larkin, and D. 
E.Khmel'nitskii, Sov. Phys. JETP {\bf 54}, 411 (1982).

\bibitem{25} O. E. Raichev and F. T. Vasko, Phys. Rev. B. {\bf 55}, 2321 
(1997).

\end{references}
\end{document}